\documentclass[11pt,letterpaper]{article}
\usepackage{amsfonts}
\usepackage{graphicx}

\begin{document}

\title{Automated deconvolution of structured mixtures from bulk tumor genomic data}
\date{}
\maketitle
\vspace*{-0.9in}
\begin{center}
\author{Theodore Roman\,$^{1}$, Lu Xie\,$^{2}$ and Russell Schwartz\,$^{1,3,*}$}
\end{center}
$^{1}$Computational Biology Dept., Carnegie Mellon University, Pittsburgh, PA, 15213, USA  \\
$^{2}$Statistics Dept., Carnegie Mellon University, Pittsburgh, PA, 15213,
USA\\
$^{3}$Dept. of Biological Sciences, Carnegie Mellon University, Pittsburgh, PA, 15213, USA\\
$^{*}$To whom correspondence should be addressed: russells@andrew.cmu.edu

\begin{abstract}
\noindent
\textbf{Motivation:} As cancer researchers have come to appreciate the importance of intratumor heterogeneity, much attention has focused on the challenges of accurately profiling heterogeneity in individual patients.  Experimental technologies for directly profiling genomes of single cells are rapidly improving, but they are still impractical for large-scale sampling.  Bulk genomic assays remain the standard for population-scale studies, but conflate the influences of mixtures of genetically distinct tumor, stromal, and infiltrating immune cells.  Many computational approaches have been developed to deconvolute these mixed samples and reconstruct the genomics of genetically homogeneous clonal subpopulations.  All such methods, however, are limited to reconstructing only coarse approximations to a few major subpopulations.   In prior work, we showed that one can improve deconvolution of genomic data by leveraging substructure in cellular mixtures through a strategy called simplicial complex inference.  This strategy, however, is also limited by the difficulty of inferring mixture structure from sparse, noisy assays.

\noindent
\textbf{Results:} We improve on past work by introducing enhancements to automate learning of substructured genomic mixtures, with specific emphasis on genome-wide copy number variation (CNV) data.  We introduce methods for dimensionality estimation to better decompose mixture model substructure; fuzzy clustering to better identify substructure in sparse, noisy data; and automated model inference methods for other key model parameters.  We show that these improvements lead to more accurate inference of cell populations and mixture proportions in simulated scenarios.  We further demonstrate their effectiveness in identifying mixture substructure in real tumor CNV data.

\noindent
\textbf{Availability:} Source code is available at

\noindent
{\tt http://www.cs.cmu.edu/\textasciitilde russells/software/WSCUnmix.zip}
\end{abstract}

\section{Introduction}
Tumor heterogeneity is now recognized as a pervasive feature of cancer biology with implications for every step of cancer development, progression, metastasis, and mortality.  Most solid tumors exhibit some form of hypermutability phenotype \cite{Loeb01}, leading to extensive genomic variability as tumor cell populations expand \cite{Marusyk12}.  Studies of single cells by fluorescence in situ hybridization (FISH) \cite{Shackney04, HeselmeyerHaddad12} have long revealed extensive cell-to-cell variability in single tumors, an observation that has since been shown by single-cell sequencing technologies to occur with a far greater scale and variety of mechanisms than previously suspected (e.g., \cite{Navin11,Ling15}).  Furthermore, studies of clonal populations across progression stages have revealed that it is often rare cell populations within this heterogeneity that underlie progression, rather than the dominant clones \cite{HeselmeyerHaddad12}.  Indeed, heterogeneity itself has been shown to be predictive of progression and patient outcomes \cite{Fisher13}.  All of these observations have suggested the importance of having ways of accurately profiling tumor heterogeneity, for both basic cancer research and translational applications.

Experimental technologies for profiling tumor heterogeneity are constantly improving, but are so far impractical for systematically profiling variability genome-wide in large patient populations.  FISH and related imaging technologies are practical for profiling many thousands of cells, but only at limited sets of preselected markers \cite{HeselmeyerHaddad12}.  Single-cell sequencing can derive genome-wide profiles of hundreds of cells in single tumors \cite{Wang10,Navin11}, but is so far cost-prohibitive for doing so in more than tiny patient populations.  Furthermore, technical challenges make it difficult to develop accurate profiles of structural variations, such as copy number variations (CNVs), which are the major drivers of progression in most solid tumors \cite{Heng13}.  Bulk regional sequencing can profile small numbers of tumor sites per patient in large patient populations \cite{Tsao00} but provides only a coarse picture of the heterogeneity within each site. 

These technical challenges to assessing heterogeneity experimentally have led to enormous interest in computational deconvolution (also known as mixed membership modeling or unmixing) methods as a way of computationally separating cell populations from mixed samples.  Originally proposed as a way of correcting for stromal contamination in genomic measurements \cite{Etzioni05}, such methods were later extended to reconstructing clonal substructure \cite{Quon09} and subclonal evolution \cite{Schwartz10} among tumor cell populations.  The past few years have seen an explosion of such methods for deconvolution of numerous forms of genomic data sources (e.g., \cite{Roth12,Su12,Oesper13,Ha14,Li14,Roman15}).  All such methods, however, are limited in accuracy and capable of resolving at best a few major clonal subpopulations, a small fraction of the heterogeneity revealed by single-cell experimental studies.  These limits result from an inherent difficulty of separating high-dimensional mixtures, especially from sparse, noisy data.  The gap between the heterogeneity we know to be present and what we can resolve by deconvolution is enormous, suggesting a need for further methodological advances.  

In prior work, we proposed that one could better resolve genomic mixtures by taking account of the fact that we would expect such mixtures to exhibit extensive substructure \cite{Roman15}.  That is, an individual tumor or tumor site is not likely to be a uniform mixture of all cell types observed across all tumors in a study.  Rather, one can expect distinct samples to group into subsets that share more or fewer cells depending on how closely related they are to one another.  All tumors can be expected to share some contamination by normal cells while tumors with common subtypes can be expected to share cell states characteristic of those subtypes.  Likewise, tumor regions might be expected to share more similarity with those nearby than with those more distant in the patient.  This kind of substructure is in principle exploitable to improve our ability to reconstruct accurate mixed membership models.  Specifically, by deconstructing tumor samples into subgroups with similar mixtures, one can decompose the problem of reconstructing a high-dimensional mixture into the easier problem of reconstructing several overlapping lower-dimensional mixtures.  We previously showed how to implement such an approach by adapting an earlier deconvolution strategy for uniform mixtures that was based on identifying geometric structures (simplices) of tumor point clouds in genomic space \cite{Schwartz10,Tolliver10}.  We subdivide these point clouds into low-dimensional subsimplices that collectively constitute a higher-level object known as a simplicial complex. This prior work used a pipeline of several sequential steps to transform a genomic point cloud into a structured mixed membership model \cite{Roman15}:
\begin{enumerate}
\item Pre-processing / filtration
\item Dimensionality reduction
\item Pre-clustering (partitioning) into uniform submixtures
\item Unmixing submixtures
\item Unifying mixtures into a structures simplicial complex model
\end{enumerate}
The resulting pipeline established a proof-of-concept for the approach, but also introduced several difficult computational challenges.  For example, it required accurately pre-specifying the number of partitions and the dimensionality of each of the partitions, both difficult inference problems in themselves that require significant knowledge of the system under study.

In the present work, we improve on this proof-of-concept method by tackling several subproblems on the path to more completely automating inference of substructured genomic mixtures from populations of tumor samples. We have eliminated several nuisance parameters from the prior work, most notably by introducing methods for automated dimensionality estimation of subsimplicies and automated maximum likelihood inference of other previously user-defined parameters.  We have further introduced a fuzzy clustering approach customized for simplicial complex decomposition to better handle uncertainty in substructure decomposition.  We verify the effectiveness of these new methods on simulated data sets of known ground truth in comparison to our prior work and  to a more conventional Gaussian mixed membership model.  We further demonstrate the practicality of the new methods and their competitiveness with related work through a study of ovarian (OV) tumor CNV data drawn from The Cancer Genome Atlas (TCGA) \cite{Weinstein13}.

\section{Methods}

In this section, we go through each step of our improved analysis pipeline, followed by a discussion of validation and application to simulated and real tumor data.
We break the full inference problem into a series of sequential steps.  Figure 1 provides a high-level overview of the process.  The following subsections provide details on each component.

\begin{figure}[t!]
\centering
\includegraphics[width=5.5in]{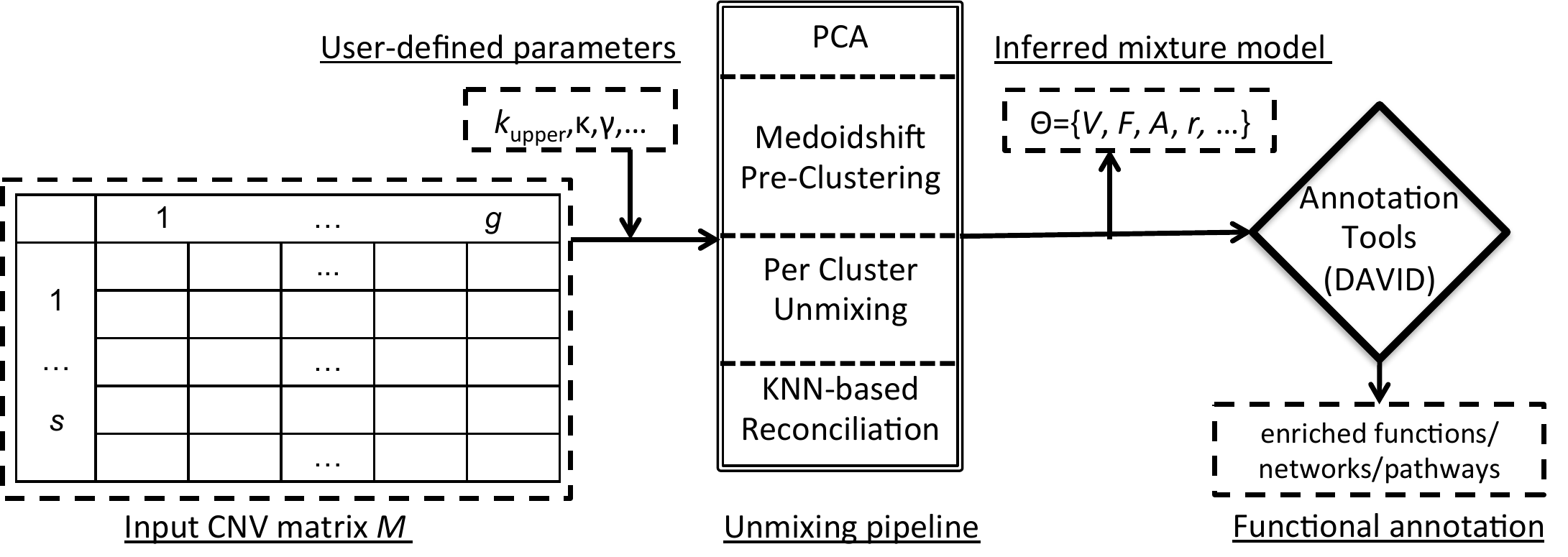}
\caption{Overview of the full analysis pipeline: Input samples are represented by collections of copy number (CN) call files, which are converted to a matrix format.  These matrix inputs are passed to our simplicial complex inference code, which infers a mixed membership model of the data and associated model likelihood.  The inference is computed by using principal components analysis (PCA), following by pre-clustering, unmixing on each substructure (simplex), and using a K-nearest-neighbor (KNN-based) reconciliation model.  The inferred mixture components are then used in downstream functional annotation to identify dysregulated pathways or term associations.}
\end{figure}

\subsection{Input and Output Data}

We conceptually model input data as a matrix $M \in \mathbb{R}^{s \times g}$, where the $s \in \mathbb{N}$ rows correspond to distinct samples (which might be biopsies of tumors in a patient population, tumor sites in a patient, or regions of a single tumor) and the $g \in \mathbb{N}$ columns correspond to probes along a genome (typically one per gene, although potentially at lower or higher resolution).  For ease of exposition, we refer to rows as tumors and columns as genes below.  We use this generic matrix format because data from many sources can be preprocessed into such a matrix (e.g., array-based data, whole-genome or whole-exome sequencing data).  We more specifically assume below that the input matrix contains CNV data as might be derived from array comparative genomic hybridization (aCGH) or DNAseq read depths.  Although the strategy proposed here might be applied to other forms of genomic measurement, CNV data would be expected to be more cleanly modeled as convex combinations of pure cell types.  
Our goal is to decompose the rows of $M$ into an approximately convex combination of a smaller set of unknown mixture components (putative cell populations).  More formally, we seek a decomposition
\begin{equation}
M = FV + \epsilon \label{eq:01}
\end{equation}
where $F \in \mathbb{R}^{s \times k}$ are mixture proportions, $V \in \mathbb{R}^{k \times g}$ are unmixed subpopulations, $k \in \mathbb{N}$ is the number of inferred cell subpopulations, and $\epsilon \in \mathbb{R}^{s \times g}$ is an error matrix.  $F$ is interpreted as the mixture fractions of the pure subpopulations, also called mixing proportions, and $V$ as the inferred genomic profiles of the pure subpopulations, also called mixture components.  This interpretation leads to natural constraints on the problem: 1) $\sum_{i} F_{ij} = 1$ for a fixed $j$ and 2) $\forall i,j: 0 \le F_{i,j} \le 1$.  Given these constraints, the formal goal of the method is to compute $F$ and $V$ given $M$, with an intermediate step of determining the mixture dimension $k$.

\subsection{Pre-processing}

To begin analysis, we first pre-process $M$ into a matrix of Z-scores:
\begin{equation}
M_z = \frac{M-\mu_M}{\sigma_M} \label{eq:02}
\end{equation}
where $\mu_M$ is a vector of the mean copy numbers of each gene across all samples, and $\sigma_M$ is a vector of the standard deviations of the copy numbers.
Next, to facilitate analysis of genomic point clouds, we reduce the dimension of the data using principal components analysis (PCA) \cite{Pearson01}.  While there are more sophisticated dimensionality reconstruction strategies available, we favor PCA as a simple, standard method that has relatively modest data needs.  We identify a total of $k_{upper}$ PCs, using the Matlab {\tt pca} routine in economy mode, where $k_{upper} \in \mathbb{N} < g$ is an upper bound on the number of cell subpopulations we will infer.  In the present work, we use $k_{upper}$ = 7, intended to be approximately an upper limit on the number of distinct mixture components a method of this class might be able to infer.  We denote the PCA scores, corresponding to amounts of each PC in each tumor, as $S_M \in \mathbb{R}^{s \times k_{upper}}$.
Lastly, we normalize the scores for each PC to a $[0,1]$ range, which is then assumed by the pre-clustering technique applied in the next section \cite{Roman16}.  We compute the 0-1 normalized version of $S_M$ as 

\begin{equation}
S_{[0,1]} = \frac{S_M - \min S_M}{\max S_M - \min S_M}\label{eq:03}
\end{equation}

\noindent
where the minimums and maximums are computed for each PC, taken over all samples.

\subsection{Pre-clustering}
We next pre-cluster data to identify initial candidate subsets of tumors inferred to draw from the same set of mixture components.  Each such subset will correspond to a distinct subsimplex of the full simplicial complex to be inferred.  While this is a clustering problem, it is a non-standard one in that we seek to cluster data into distinct low-dimensional subspaces of a contiguous higher-dimensional point cloud, rather than into disjoint subclouds as is in conventional clustering.  We recently developed a specialized clustering method for this purpose \cite{Roman16}, based on a two-stage variant of medoidshift clustering \cite{Sheikh07}.  We initially cluster in Euclidean PC space to reduce the raw data to a smaller set of representative data points.  We then cluster these representatives under a negative-weight exponential kernel function using the ISOMAP distance measure \cite{Tenenbaum00}, a form of geodesic metric measuring distance between data points through a k-nearest-neighbor graph of the input point cloud. The combination of ISOMAP distance and negative exponential kernel produces a clustering in which cluster centroids are approximately extremal points of the simplicial complex that serve to pull apart distinct subspaces of the point cloud.  The initial Euclidean clustering suppresses noise, which otherwise makes the negative exponential kernel highly sensitive to outlier data. We refer the reader to \cite{Roman16} for full details.  
At the end of this process, we are left with a small set of cluster representatives

\begin{equation}
M_{2 stage} = { \cup_i M_{N(x_i)} }, \label{eq:04} 
\end{equation}

\noindent
where each representative is itself a point in $S_{[0,1]}$, and a corresponding clustering of all tumors $C=\{C_1, \ldots, C_r\}$, $S_{[0,1]}=\bigcup_{C_i \in C} C_i$.

We further assess uncertainty of the cluster assignments by determining a relative statistical weight of each data point in each cluster.  We use a weight function based on a folded multivariate normal distribution, where the mean of the function is a 0 vector, the covariance matrix is the identity multiplied by the distance from each cluster center to the mean of all cluster centers, and the value at which the density function is evaluated is the distance from $x_i$ to $C_j$ in ISOMAP space.  After these relative weights have been derived, we convert them to probabilities of assignment of each point to each cluster.  If we denote the raw weight of the $i^{th}$ data point as a vector $R_i$, then we can define the normalized weight vector:
\begin{equation}
W_{i} = \frac{R_i - \min_{C_j \in C} R_i}{\max_{C_j \in C} R_i - \min_{C_j \in C} R_i} \label{eq:05}
\end{equation}
The clustering in principle depends on a chosen neighborhood size, although a scan over all possible neighborhood sizes found no sensitivity of the final model likelihood to this parameter. 

\subsection{Dimensionality Inference}

We next seek to estimate the dimension of each cluster, which will correspond to the number of mixture components inferred for that cluster.  The major challenge of this step is distinguishing a genuine axis of variation from random noise, particularly when working with sparse, noisy genomic measurements.  Intuitively, we identify dimension by iteratively adding axes of variation via PCA until we can no longer reject the hypothesis that variance in the next dimension is distinguishable from noise.  

We first build a model of expected noise per dimension by randomly sampling data points of pure Gaussian noise with mean 0 and identity covariance.  We then perform PCA on this random point cloud and estimate the mean $\mu_{G(i)}$ and standard deviation $\sigma_{G(i)}$ of the point cloud for each PC $i \in {1,...,k_{upper}}$.  We then identify the smallest $i \leq k_{upper}$ such that the standard deviation of the true data in PC $i$ is smaller than $\mu_{G(i)} + \kappa \sigma_{G(i)}$, where $\kappa$ defines a significance threshold in standard deviations.  In the present work, we set $\kappa=3$ to yield effectively a significance threshold of $< 0.01$ for rejecting the hypothesis that the next dimension can be explained by Gaussian noise.  The result of this module, then, is a vector of inferred dimensions of each of the clusters:
$D \in \{1,\ldots,k_{upper}\}^{r}$.
We would expect this test to be conservative (underestimate true dimension), although less so as the size of the data set and its precision increases.  We found it necessary to use a custom-made conservative dimensionality estimator, as opposed to a more standard technique (e.g., \cite{Cheng09}), because the number of data points available in this application is much smaller than is typically assumed by methods in this problem domain. 

\subsection{Cluster-wise Unmixing} 

We next seek to establish an initial mixed membership model by separately unmixing each cluster, using the inferred dimension from the previous step as the number of mixture components.  We establish the model by minimizing an objective function based on the noise-tolerant geometric unmixing method of \cite{Tolliver10}:
\begin{equation}
P(\theta | X) \propto \prod_{i=1}^{r}(exp(-\sum_{j=1}^{s}(|x_i -F_{j}^{i}V_{j}^{i}|W_{j}^{i}))MST(V_{j},A_{j})^{\gamma}\beta)\label{eq:06}
\end{equation}
where $\gamma$ is a regularization penalty set based on an estimated signal-to-noise ratio (SNR) of the data source~\cite{Roman15}, $V$ are the inferred vertices, $A$ is the adjacency matrix, $MST$ is a minimum spanning tree cost, $W$ is the relative weight function computed above, $F$ are the inferred mixture components, $x_i$ is the $i^{th}$ data point, $\beta$ is a BIC penalty for model complexity \cite{Schwarz78} and $| |$ is L1 distance.  The first term penalizes data points outside the bounding simplex via an exponentially-weighted L1 penalty.  The MST term captures a form of parsimony model on the simplex itself intended to penalize the amount of mutation from a common source needed to explain the simplex vertices (mixture components) \cite{Roman15}.  We optimize for the objective function via the Matlab {\tt fmincon} function, fitting $V$ and $F$ to assign mixture components and mixture fractions to each cluster independently.

\subsection{Reconciliation of Subsimplices into a Simplicial Complex}

We next seek to join the discrete simplices, each modeling a subset of tumors as a uniform mixture, into a unified simplicial complex.  We accomplish this by merging simplex vertices if we cannot reject the hypothesis that they represent distinct points in genomic space.
We first establish a probability model using the k-nearest-neighbors graph on tumors and vertices by modeling the set of overlapping neighbors between two vertices via a hypergeometric distribution.  On the assumption two vertices draw their neighbor sets independently from the pool of all tumors, the expected number of data points in common would be
\begin{equation}
\frac{|N_1||N_2|}{N}\label{eq:07}
\end{equation}

\noindent
where there are $N$ data points, $|N_1|$ nearest neighbors of the first vertex, and $|N_2|$ neighbors of the second vertex.  We merge two vertices when the number of observed overlapping nearest neighbors is above expectation.  We empirically determined on our synthetic data that the method is insensitive to the number of nearest neighbors for choices between 2 and $\sqrt{N}$ and chose $k=15$ nearest neighbors arbitrarily within this range for the real data.  This approach replaces computationally costly bootstrap estimates used in our prior work \cite{Roman15}.

\subsection{Validation on Synthetic Data}

To test our methods, we first sought to determine their effectiveness in reconstructing a series of mixed membership models of known ground truth.  For this purpose, we defined seven basic simplicial complex structures of varying dimension and geometries, as in our prior work \cite{Roman15}.  These seven models are designed to encompass a wide variety of evolutionary scenarios and combinations of subsimplex dimensions achieveable with a small number of mixture components.  
The simplicial models and corresponding tumor evolution scenarios are illustrated in Fig.~2. 	

\begin{figure}[t!]
\centering
\includegraphics[width=5.5in]{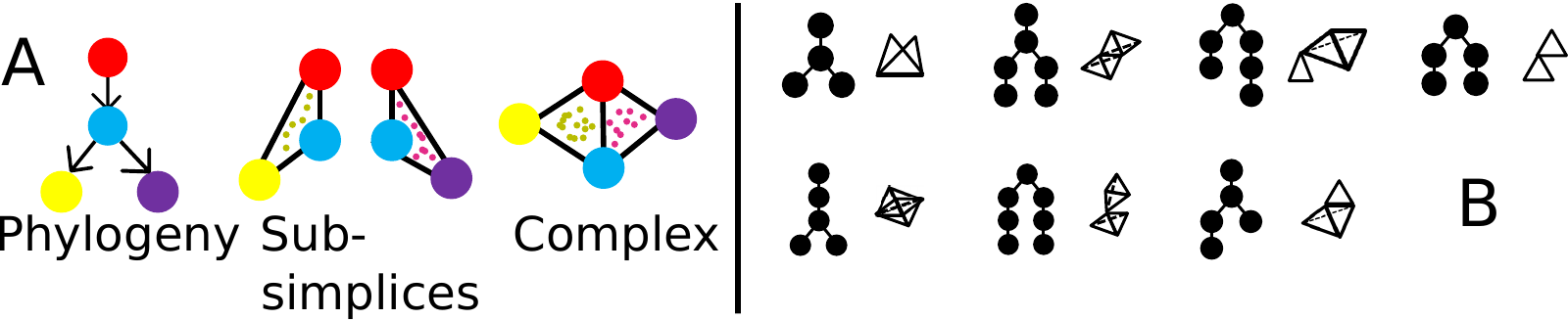}
\caption{The simplicial complex inference problem. (A) Illustration of a tumor phylogeny describing progression of a set of cell types, a realization of a population of tumors drawn from distinct lineages in the phylogeny as a pair of low-dimensional subsimplices, and a model of the resulting full simplicial complex as a union of simplices across lineages joined by their shared cell types. (B) Illustration of seven scenarios of tumor phylogeny (left), each with a corresponding simplicial structure (right), which are used to generate synthetic test cases for validation.}
\end{figure}

To simulate a tumor from a mixture scenario, we draw a set of vertices uniformly at random from among the subsimplices.  For each vertex, we sample a raw statistical weight uniformly at random between 0 and 1 and normalize weights across vertices to yield a uniformly random mixture of vertices.  We further add Gaussian noise to each dimension, with standard deviation varied from 0 to 1.0 in units of 0.1.  For each replicate of each model, we generate 250 tumors embedded in 50 dimensions, to yield a similar dataset after PCA to the 472 tumor samples in the real data we examine.  We perform replicates of the unmixing procedure for each scenario and noise level, varying the number of nearest neighbors from 1 to $\sqrt{250}$.  We also repeat the evaluation for neighborhood sizes corresponding to each possible unique integer distance.

We apply our deconvolution method to each such data set and assess inference accuracy by the root mean squared deviation (RMSD) of mixture components between real and inferred mixtures, normalized by the product of the number of ambient dimensions and the number of ground truth components.  This measure evaluates the fractional error in the inferred genotypes of the unmixed subpopulations.  We compare this measure between our proposed method, an earlier version of the method without fuzzy clustering, and three variants of Gaussian mixed membership modeling: one also using our automated dimension estimates, one supplied with the true total dimension of each data set, and another given only an upper bound on the dimensions.  Unlike with our previous approaches \cite{Schwartz10,Tolliver10,Roman15}, there is no straightforward way to evaluate the mixture proportions derived by the method in comparison to a ground truth, as the mixture fractions are re-weighted inside the simplicial complex objective function.

\subsection{Application to Real Tumor CNV Data}

We next demonstrate the method on ovarian (OV) CNV data from the Cancer Genome Atlas (TCGA)\cite{Weinstein13}.  We downloaded level 3 CNV data for ovarian (OV) cancer, accessed on 15 July, 2015, consisting of 472 samples.  We internally represent the data as maximal blocks of fixed copy number along the genome, as in our previous work \cite{Roman16}.  Although the blocks are potentially more or less granular than genes, our method is agnostic to the granularity of data.  From the inferred geometric structure and the loading matrix from the principal components analysis, we obtain chromosome coordinates for the CN data for significantly dysregulated regions of each inferred subtype.  We ran the full unmixing pipeline using the following parameters: maximum number of dimensions 7; number of bootstrapped replicates for pre-clustering 1000; neighborhood size for pre-clustering 1; number of nearest neighbors for vertex merger 15; cutoff for dimensionality estimation 3 standard deviations; maximum number of iterations of {\tt fmincon} per simplex: 1000.  The choice of dimensionality reflects an approximate upper bound on the number of dimensions methods of this class can infer, and some upper bounded is needed to limit runtime.  The number of bootstrap replicates was chosen empiricaly to be sufficiently large to ensure stable bootstrapped clustering.  The maximum number {\tt fmincon} of iterations was also chosen empirically to be sufficient to reach convergence.  The neighborhood size of 1 corresponds well to assumptions made about the medoidshift clustering procedure.  For full details, see \cite{Roman16}.  The number of nearest neighbors was chosen based on tests on the simulated data showing the method to be insensitive to this parameter up to approximately $\sqrt{N}$ neighbors.  The 3 standard deviations above the mean corresponds to a p-value of $0.01$.  The runtime of the experiments depends largely upon the dimensionality of the clusters (i.e., the number of subpopulations in the tumor dataset) and the number of iterations in the minimization phase (iterations of {\tt fmincon}).

\section{Results}

\subsection{Validation on Synthetic Data}
We assess the performance of the algorithm in comparison to a prior version of simplicial complex unixing with medoid-shift clustering although without weighting and to a more generic Gaussian mixture model (GMM).  To test the value of the dimension estimation aspect of the present work specifically, we compare to versions of the GMM fed the true dimension, an upper bound on the dimension, or using dimension inferred by the same method as our simplicial complex approach.  Table 1 shows he results for each of the seven scenarios.

\begin{table}[t!]
\caption{Error in mixture estimation.  Each column gives the mixture component inference error in normalized RMSD between true components and the best-matching inferred component for each.  For the two methods using dimension estimation, $GMM^{2}$ and $SC^{wt.}$, we also provide the difference between the number of true and inferred components in parentheses.  The columns correspond to GMM with an upper bound on number of components ($GMM^{1}$), GMM with automated dimension estimation ($GMM^{2}$), GMM given the true number of components ($GMM^{3}$), unweighted simplicial complex with an upper bound on components ($SC^{unwt.}$), and simplicial complex with automated dimension estimation ($SC^{wt.}$, i.e., the present method).}
\centering
\begin{tabular}[t]{llllll}
Scenario & $GMM^{1}$ & $GMM^{2}$ & $GMM^{3}$ & $SC^{unwt.}$ & $SC^{wt.}$\\
\hline
1 & 0.0513 & 0.0133 (-3) & 0.0267 & 0.0253 & 0.0188 (0)\\
2 & 0.0469 & 0.0133 (-3) & 0.0243 & 0.0256 & 0.0185 (+8)\\
3 & 0.0534 & 0.0150 (-6) & 0.0367 & 0.0267 & 0.0197 (-1)\\
4 & 0.0446 & 0.0150 (-5) & 0.0319 & 0.0264 & 0.0197 (+6)\\
5 & 0.0482 & 0.0150 (-4) & 0.0224 & 0.0263 & 0.0124 (+1)\\
6 & 0.0559 & 0.0143 (-4) & 0.0262 & 0.0261 & 0.0160 (-1)\\
7 & 0.0590 & 0.0143 (-5) & 0.0297 & 0.0256 & 0.0146 (-2)\\\hline
\end{tabular}
\end{table}

Comparing the two simplicial complex methods suggests that the novel methods introduced in the present work do significantly improve accuracy relative to our prior strategy of upper-bounding dimension in subsimplices.  This is true even though the dimension estimates can be inaccurate, with a tendency either to slightly underpredict or significantly overpredict the true dimension (mean error $+1.6$).  The GMM methods have mixed performance depending on the number of components inferred.  Always using the upper-bound yields poor performance, which is improved by feeding the method the true number of components.  Suprisingly, GMM with inferred components yields generally the lowest RMSDs despite generally significantly underestimating the dimension of the data (mean error $-4.3$).  This seemingly counterintuitive observation reflects an artifact of the RMSD measure, in which one can get good RMSD scores by merging multiple true components into a single inferred component, effectively fitting the components well but losing the desired mixture structure.  The approach of subdividing the problem into smaller, low-dimensional subproblems does allow substantially better estimation of the qualitative simplicial structure, but more sophisticated probabilistic models of these substructures may lead to better quality decompositions than our current geometry-based unmixing approach.

\subsection{Application to Ovarian Tumor CNV Data}

To better illustrate the utility of the approach on real cancer data, we next applied it to a dataset of CNV data from OV tumors taken from the TCGA \cite{TCGAovarian2011}.  Using parameters described in Section 2.8, the weighted unmixing procedure inferred a simplicial structure consisting of five vertices, with substructures of a tetrahedron and a triangle conjoined at a line.  The results are illustrated in Fig.~3, which shows the true point cloud and inferred simplicial structure projected into PC space.  The figure shows two views, capturing the first three and next three PCs.  The first view clearly highlights a simplicial complex structure apparent in the raw data, consisting three arms branching from a common point.  The second view shows some of the variability in high dimensions, which can be presumed to explain why a more complex mixture structure is inferred than seems apparent in the first three PCs.

\begin{figure}[t!]
\centering
\includegraphics[width=5.5in]{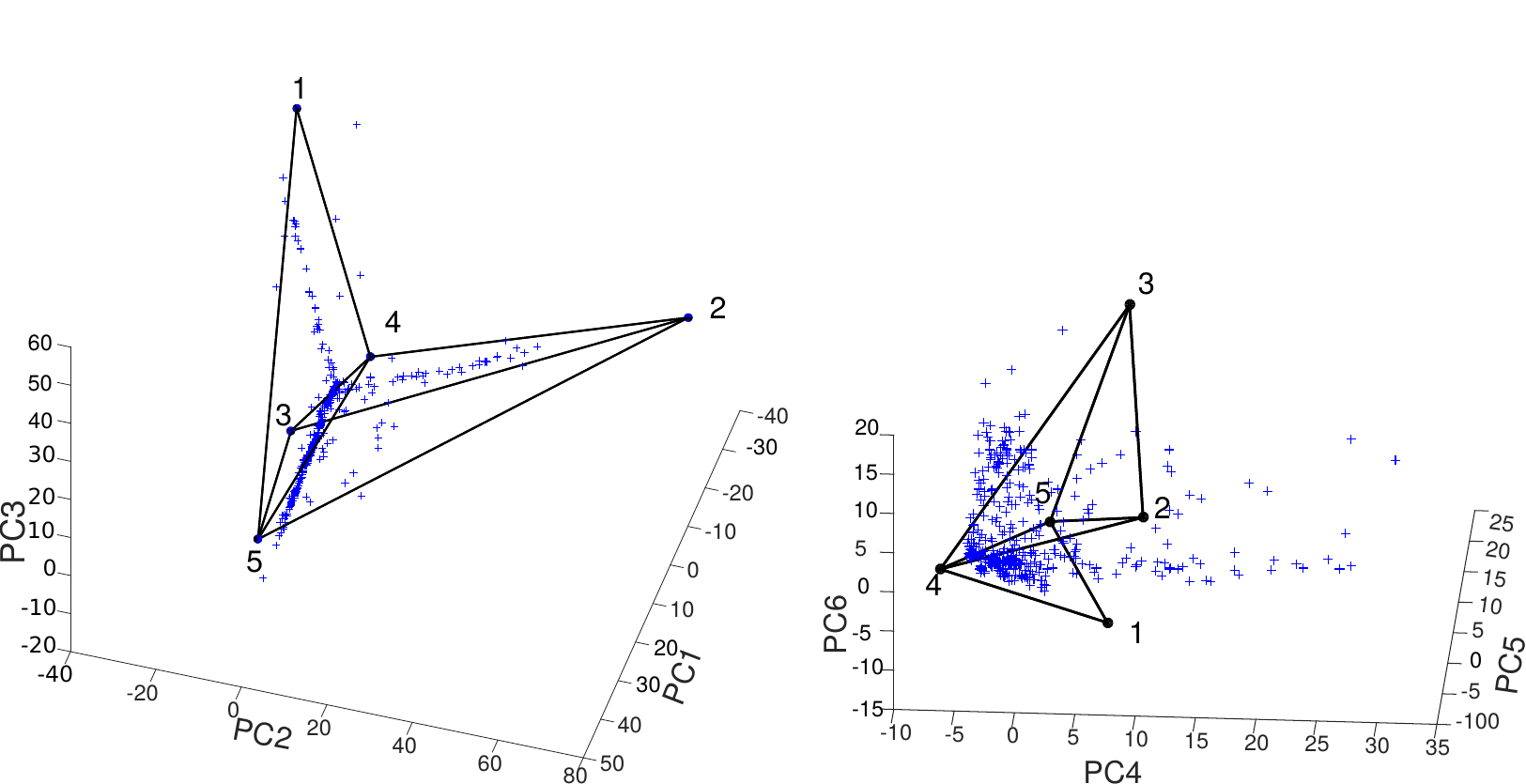}
\caption{Visualization of the Ovarian data (blue) and inferred components (black), with inferred connections in black line.
We can interpret the geometry in terms of a phylogeny of tumors, with the two subsimplices, a tetrahedron and a triangle joined at an edge, corresponding to a putative high-level split into two subtypes from a common normal ancestor (vertex 4) and a putative early state of progression common to all subtypes (vertex 5).  From there, two evolutionary trajectories emerge, where one adds a leaf node at vertex 1, while the other contains vertices 2 and 3.  This phylogenetic intepretation can be derived from a minimum spanning tree among vertices as in \cite{Roman15}, with the edge lengths of this tree model providing one component of the model likelihood function.}
\end{figure}

To assess functional significance of inferred mixture components, we projected them from PC back to genomic space and identified those CNV blocks with copy number above a Z-score of 2 in each component at a resolution of 10 kb.  We then identified genes overlapping these elevated blocks using the UCSC Genome Browser \cite{Kent02,Karolchik04} to obtain a list of genes associated with the elevated blocks.   We identified statistically significant term associations for the resulting gene lists for each mixture component using the DAVID toolkit \cite{Dennis03}.  Table 2 provides significant term associations at p-value $<0.01$ as clustered by DAVID.  Most term associations were associated with vertices 1 and 5, with particular relevance for terms associated with chromatin structure.  Vertex 2 which distinguishes the triangular simplex from the tetrahedral, provided significant association to terms associated with regulation by zinc finger transcription factors.  Vertex 4, which we identify as an inference of normal stromal cells, shows no term associations.  There are also none found for vertex 3, which may indicate that it reflects only a minor axis of variation in these data.  In addition to the above pathway results, DAVID also found weakly significant ($p < 0.05$) tissue-specific expression for the highlighted genes in multiple inferred vertices for uterine and ovarian carcinoma tissue, as well as for lung tissue. 

\begin{table}[ht!]
\caption{DAVID results for enriched term associations of OV simplicial complex vertices.  Each row identifies the relevant vertex, term association, DAVID cluster of the term, and p-value. $^*$: terms condensed, reporting highest corrected p-value.}
\centering
\begin{tabular}[t]{llll}
Vertex & Term & Cluster & p-value\\
\hline\hline
1 & GO:nucleosome*& 1 & $\le1.53 \times 10^{-29}$\\
1 & GO:protein-DNA complex*& 1 &$\le8.48 \times 10^{-30}$\\
1 & GO:Chromatin*& 1 &$\le6.81 \times 10^{-10}$\\
1 & GO:DNA Packing/Binding & 1 &$\le 2.76 \times 10^{-13}$\\
1 & KEGG:lupis erythematosis & 1 & $2.31 \times 10^{-19}$\\
1 & Pather:Histone& 1 &$1.17 \times 10^{-18}$\\
1 & Reactome:Telomere Maint.& 1 &$1.08 \times 10^{-18}$\\
1 & GO:Cellular Macromolecular component org.*& 1 &$9.20 \times 10^{-4}$\\
5 & GO:nucleosome*& 1 &$\le 1.53 \times 10^{-29}$\\
5 & GO:protein-DNA complex*& 1 &$\le 8.48 \times 10^{-30}$\\
5 & GO:Chromatin*& 1 &$\le 6.81 \times 10^{-10}$\\
5 & GO:DNA Packing/Binding& 1 &$2.76 \times 10^{-13}$\\
5 & KEGG:lupis erythematosis& 1 &$2.31 \times 10^{-19}$\\
5 & Panther:Histone & 1 & $1.17 \times 10^{-18}$\\
5 & Reactome:Telomere Maint. & 1 & $1.08 \times 10^{-18}$\\
5 & GO:Cellular Macromolecular component org.* & 1 & $9.20 \times 10^{-4}$\\
\hline
1 & GO:DNA Binding*& 2 &$\le 3.19 \times 10^{-3}$\\
2 & Panther:KRAB box TF & 2 & $6.96 \times 10^{-7}$\\
2 & Panther:ZF TF & 2 & $8.63 \times 10^{-4}$\\
5 & GO:DNA Binding* & 2 & $\le 3.19 \times 10^{-3}$\\
\hline
1 & Panther: Lactation, mammary development & 3 & $3.63 \times 10^{-7}$\\
5 & Panther:Lactation, mammary development & 3 & 3$.63 \times 10^{-7}$\\
\hline
1 & GO/KEGG: Olfactory* & 4/5 & $\le 5.29 \times 10^{-3}$\\
5 & GO/KEGG: Olfactory* & 4/5 & $\le 5.29 \times 10^{-3}$\\
\hline
1 & Panther:MHCI-mediated immunity & 6 & $6.63 \times 10^{-3}$\\
5 & Panther:MHCI-mediated immunity & 6 & $6.63 \times 10^{-3}$\\
\hline
1 & Panther:KRAB box TF & 7 & $1.24 \times 10^{-3}$\\
5 & Panther:KRAB box TF & 7 & $1.24 \times 10^{-3}$\\
\hline
5 & GO/Panther: Phosphate* & 8 & $\le 5.49 \times 10^{-3}$\\
\hline
\end{tabular}
\end{table}

Direct comparison of the present method with other standard genomic deconvolution approaches is challenging due to differing forms of input data required by all leading alternatives.  Nonetheless, we can compare our predictions to others performed on some of the same tumors for the purpose of purity estimation \cite{Aran15}.  Purity estimation is not the purpose of our method, but it can be considered a subproblem of full mixture decomposition.  To derive a comparable measure of tumor purity, we identified vertex 4 of our simplicial complex model, for which no elevated regions of the genome were detected, as the most plausible representative of genetically normal contaminating stromal cells.  We then estimated the purity of each tumor by a sum of its inferred fraction of non-normal cells over subsimplices, weighted by its inferred probability of belonging to the given subsimplex   Our estimates compared to those of Aran et al.~\cite{Aran15} have a Spearman correlation coefficient of 0.1436, giving a weakly significant p-value of $0.0274$ and showing that our decomposition is concordant with prevailing purity estimates on a large set of real OV tumors.   Table 3 provides a confusion matrix comparing our method to THetA \cite{Oesper13}, a leading alternative method, in their concordance with the TCGA purity estimation pipeline \cite{Aran15}.  The Aran et al.~consensus purity estimates are derived from a combination of ABSOLUTE \cite{carter2012absolute} (which uses genomic and karotyping information), annotation based on immunological staining, and additional measures for which calls are unavailable to us.  Treating their consensus as our best estimate of a ground truth, we observe a weakly significant correlation to our estimates although an insignificant anti-correlation to THetA.  Since ABSOLUTE and the staining are components of this consensus call, their high concordance is to be expected.  ABSOLUTE shows a significant correlation with the staining data, which neither our method nor THetA shows.  This may suggest the ABSOLUTE method is superior to our own at this task, but could also reflect the fact that ABSOLUTE has direct karyotyping information as input, which neither our method nor THetA does, or the fact that ABSOLUTE is designed for purity estimation while neither our method nor THetA are.  We thus hesitate to draw strong conclusions from these data given the wide differences in goals and input data types between methods, but nonetheless suggest they do show that our method is at least competitive with a popular alternative genomic deconvolution method at the task of separating tumor and non-tumor subpopulations.

\begin{table}[h!]
\caption{Confusion matrix of correlations between purity estimates on TCGA OV data for WSC (our method), THetA, ABSOLUTE, expert annotation from immunological staining, and a consensus call derived from ABSOLUTE, staining, and additional estimators for which calls are unavailable.  Each entry presents the correlation coefficient between shared samples OV followed by statistical significance in parentheses.}
\centering
\begin{tabular}[t]{cccccc}
& WSC & THetA & ABSOLUTE & Staining & Consensus\\
\hline
WSC & 1 (0) & -0.11 (0.11) & 0.035 (0.60) & -0.035 (0.59) & 0.14 (0.027)\\
THetA & -0.11 (0.11) & 1 (0) & -0.041 (0.68) & -0.13 (0.086) & 0.034 (0.79)\\
ABSOLUTE & 0.035 (0.60) & -0.041 (0.68) & 1 (0) & 0.32 (2.2e-14) & 0.85 (1.9e-152)\\
Staining & -0.035 (0.59) & -0.13 (0.086) & 0.32 (2.2e-14) & 1 (0) & 0.27 (4.2e-11)\\
Consensus & 0.14 (0.027) & -0.030 (0.70) & 0.85 (1.9e-152) & 0.27 (4.2e-11) & 1 (0)\\
\hline
\end{tabular}
\end{table}

\section{Conclusions}
We have developed novel strategies for taking better advantage of mixture substructure in deconvolution of mixed genomic data from heterogeneous tumor samples.  This contribution is intended to advance a theoretical method for better taking advantage of substructure in complex genomic mixtures, an approach that might be incorporated into many existing approaches for cell type deconvolution using assorted data types or inference models.  The advances in the present paper bring us closer to the goal of deriving precise models of mixture substructure in the face of sparse, noisy genomic data without the need for extensive expert intervention. For this purpose, we have introduced new strategies for automated inference of subcluster dimensions, automated construction of a global simplicial complex structure, and better deconvolution of submixtures on small samples with uncertain subclustering.  We have shown that we can automatically learn model structure from realistic sizes of data set without degrading performance of the model relative to methods requiring significantly more user intervention.  While we focus specifically on deconvolution of CNV data, we believe that this simplicial complex approach for substructured mixture decomposition could be adopted for a variety of genomic data types (CNV, SNV, RNA, methylation) and technologies (array, sequence).

This ultimate goal of the present work is to make sophisticated mixture deconvolution approaches more widely accessible to a non-expert community, by allowing them to be incorporated more broadly into a variety of deconvolution approaches in the literature.  Much work still remains, though, both in better automating approaches and improving inference quality.  There are still several (hyper-)parameters for which the task of automated learning remains challenging.  While automated dimension estimation appears valuable in improving simplicial complex models, deriving accurate estimates is a significant challenge for noisy data \cite{Choi04}.  Integration of multiple forms of genomic data into a common mixture framework is likewise a promising but challenging direction for improving inference quality.  In particular, as single-cell methods become more cost-effective, combinations of bulk and single-cell data may prove particularly informative.  Finally, the simplicial complex models themselves require refinement to better capture the real sources of genomic mixture substructure they are meant to model, including substructure imposed by common pathways of subtype evolution, spatial constraints in the tumor microenvironment, and other sources of mixture substructure that do not conform well to our current simplicial complex model.  

\section*{Acknowledgments}

We thank lab members Marcus A. Thomas and Alan Shteyman for their helpful discussions and input for the manuscript.

\section*{Funding}

Portions of this work have been funded by the U.S. National Institutes of Health via awards 1R01CA140214 (RS and TR), 1R01AI076318 (RS), and T32EB009403 (TR) by a Carnegie Mellon University GuSH grant (TR), the Carnegie Mellon Computational Biology Dept. (LX), and by a Commonwealth of Pennsylvania CURE grant (RS and TR).\\
\textit{Conflict of interest:} none declared.

\bibliographystyle{abbrv}
\bibliography{RomanEtAl}

\end{document}